\documentstyle[12pt]{article}
\input{epsf.sty}
 
\def\b{\beta}

\def\d{\delta}
\def\e{\epsilon}   

\def\l{\lambda}
\def\m{\mu}

\def\o{\omega}
\def\r{\rho}      
\def\s{\sigma}    

\def\x{\xi}

\def\D{\Delta}

\def\O{\Omega}

\def\mh{\hat{\m}}
\def\ph{\hat{p}}

\newcommand{\ncm}{\newcommand}
\ncm{\rencm}{\renewcommand}
\ncm{\dsp}{\displaystyle}
\ncm{\nn}{\nonumber}
\ncm{\nnn}{\nonumber\\}
\ncm{\nit}{\noindent}
\ncm{\del}{\partial}
\ncm{\av}[1]{\mbox{$\langle #1 \rangle$}}
\ncm{\avc}[1]{\mbox{$\langle #1 \rangle_{\psi}$}}
\ncm{\half}{\mbox{{\small $\frac{1}{2}$}} }
\ncm{\quart}{\mbox{{\small $\frac{1}{4}$}} }
\ncm{\tq}{\mbox{{\small $\frac{3}{4}$}} }
\ncm{\third}{\mbox{{\small $\frac{1}{3}$}} }
\ncm{\sixth}{\mbox{{\small $\frac{1}{6}$}} }
\ncm{\eigth}{\mbox{{\small $\frac{1}{8}$}} }
\ncm{\thrhalf}{\mbox{{\small $\frac{3}{2}$}} }
\ncm{\thrfor}{\mbox{{\small $\frac{3}{4}$}} }
\ncm{\twothi}{\mbox{{\small $\frac{2}{3}$}} }
\ncm{\fivtwo}{\mbox{{\small $\frac{5}{2}$}} }
\ncm{\dx}{\mbox{$\partial_{x}$}}
\ncm{\dt}{\mbox{$\partial_{t}$}}
\ncm{\dtt}{\mbox{$\partial_{t}^2$}}
\ncm{\un}{1\!\!1}
\ncm{\RE}{\mbox{Re}}
\ncm{\IM}{\mbox{Im}}
\ncm{\Tr}{\mbox{tr}\,}
\ncm{\diag}{\mbox{diag}\,}
\ncm{\Det}{\mbox{Det}\,}
\ncm{\ra}{\rightarrow}
\ncm{\la}{\leftarrow}
\ncm{\dg}{\dagger}
\ncm{\pr}{\prime}
\ncm{\ha}{\hat{a}}
\ncm{\hP}{\hat{P}}
\ncm{\sL}{\sqrt{\Lambda}}
\ncm{\lb}{\overline{\lambda}}
\ncm{\aplt}{ \mbox{}_{\textstyle \sim}^{\textstyle < }     }
\ncm{\apgt}{ \mbox{}_{\textstyle \sim}^{\textstyle > }     }
\ncm{\Oa}{\mbox{$\mbox{O}(a)$}}
\ncm{\Sp}{\hspace{1.0cm}}
\ncm{\capit}[1]{\caption{\it #1}}
\def\be{\begin{equation}}
\def\ee{\end{equation}}
\def\bea{\begin{eqnarray}}
\def\eea{\end{eqnarray}}
\rencm{\thefootnote}{\mbox{\protect{$\fnsymbol{footnote}$}} }
\ncm{\front}[5]{
\begin{titlepage}
\noindent {#1} \hfill {#2}\\
\begin{center}
\vspace{1.5\baselineskip}
{\Large\bf  #3  } \\
\vspace{2\baselineskip}
\vspace{1.5\baselineskip}
 #4\\
\vspace{1.5\baselineskip}

University of California at San Diego,\\
Department of Physics 0319,
La Jolla, CA 92093, USA.
 
\end{center}
\vfill
{\bf Abstract}\\
 #5
\end{titlepage} }
 
\begin{document}
\front{June 1994}{UCSD/PTH 94-3}
{Investigation of Laplacian Gauge Fixing \\ for U(1) and SU(2)
Gauge fields }
{Jeroen C. Vink\footnote{e-mail: vink@yukawa.ucsd.edu} }
{
The Laplacian gauge on the lattice is investigated  numerically
using U(1) and SU(2) gauge fields. 
The problem of Gribov ambiguities is addressed and to asses the smoothness of 
the gauge fixed configurations, they are compared to configurations fixed to
the Landau gauge. The results of these comparisons with the Landau
gauge indicate that Laplacian gauge fixing works very well in practice
and offers a viable alternative to Landau gauge fixing.
}
 
\section{Introduction}
 
Gauge fixing, in particular gauge fixing to the smooth Landau gauge,
has found several applications in numerical simulations of lattice
gauge theories. Gauge fixing is unavoidable when investigating gauge
variant quantities, such as quark or gluon propagators \cite{MaOg87}.
Gauge fixing is also very useful to construct improved, extended,
versions of gauge invariant operators, in which the strings of
link fields that make these operators gauge invariant are omitted.
Leaving out these $U$-fields greatly improves the signal to noise
ratio for these operators \cite{WeakMa}. Finally,
in the `Rome approach' \cite{Rome} to chiral gauge theories the chiral fermion
determinant, which is gauge variant, must be computed on a gauge fixed
gauge field.
However, Landau gauge fixing suffers from Gribov ambiguities
\cite{Grib62}, also in its lattice implementation (see for example
\cite{GriLa} for recent work).
Therefore it is desirable to use a Landau-like gauge fixing prescription
which is free from such ambiguities. 

Some time ago a `Laplacian' gauge fixing prescription was proposed, which was
claimed to be free from Gribov ambiguities, similarly smooth as the
Landau gauge and relatively easy to compute in practice \cite{ViWi92}.
Recently, a practical non-perturbative implementation of the `Rome
approach' to chiral gauge theories was proposed that uses this Laplacian
gauge \cite{Vink93}.
Unlike the Landau gauge, the Laplacian gauge has not been used in 
practice and it is important to investigate its properties.
In this paper we apply Laplacian gauge fixing to (compact) U(1) gauge fields
in two dimensions and SU(2) gauge fields in four dimensions.
We shall address the problem of Gribov ambiguities of this gauge
condition and compare the smoothness of the gauge fixed configurations
with those fixed to the Landau gauge.

The paper is organized as follows. First we review the definition of
the Laplacian gauge for U(1) and SU(2) gauge fields in sect. 2. We
discuss the Gribov ambiguity of this gauge in sect. 3. 
In sect. 4 we compare the Laplacian
gauge with the Landau gauge by computing the average link and the
Fourier modes of the gauge fixed gauge fields.
Sect. 5 contains our conclusions.

\section{Laplacian gauge fixing and Gribov ambiguities}

The Laplacian gauge introduced in ref.  \cite{ViWi92},
uses eigenfunctions of the covariant Laplacian,
\be
 \sum_y \D(U)_{xy} f_y^s =  \sum_{\m} \left(2f^s_x 
          -U_{\m x} f^s_{x+\mh} - U_{\m x-\mh}^\dg f^s_{x-\mh} \right) = 
              \l^s f^s_x.   \label{LAPL}
\ee
Here we have suppressed the gauge field indices on $U$ and
we specialize to gauge groups G=U(1) or SU(2); 
ref. \cite{ViWi92} gives a more general discussion.
The gauge transformation $\O$ that defines the Laplacian gauge for G=U(1)
is computed from the eigenfunction $f^0$ with the smallest eigenvalue,
\be
 \O_x = f^{0*}_x\r^{-1}_x,  \;\; \r_x = |f^{0}_x|.
   \label{DEFOMU}
\ee

For G=SU(2) the eigenvalues have a twofold degeneracy, due to the
charge conjugation symmetry $U = \s_2U^*\s_2$, 
\be
 f^s \ra \s_2 f^{s*}.
\ee
The $\s_k$ are the usual Pauli matrices. The two degenerate
eigenfunctions with the smallest eigenvalue, $f^0$ and $\s_2f^{0*}$, define 
a $2\times 2$ matrix on all sites $x$, which is projected on SU(2)
to obtain the gauge transformation $\O_x$,
\be
  \O_x =  \r_x^{-1}i^{1/2}
              \left( \begin{array}{cc}
                      f^{0*}_{1,x} & f^{0*}_{2,x} \\
                      if^0_{2,x} &  -if^0_{1,x} \end{array} \right),\;\;
     \r_x = (|f^0_{1,x}|^2 + |f^0_{2,x}|^2)^{1/2}       
   \label{DEFOMSU}
\ee
The two degenerate eigenfunctions $f^0$ and $\s_2f^{0*}$ are normalized,
$\sum_x (|f^0_{1,x}|^2 + |f^0_{2,x}|^2)= 1$, and orthogonal.
Both for G=U(1) and SU(2) the $\r$ is a real number; for other gauge
groups it is a positive definite hermitian matrix \cite{ViWi92}.

As discussed in more detail in ref. \cite{ViWi92} the prescriptions 
(\ref{DEFOMU}) and (\ref{DEFOMSU}) for
$\O$ are  unambiguous, except when either
the lowest eigenvalue is degenerate for U(1) or more than two-fold
degenerate for SU(2), because in that case it is not clear how to choose
the eigenfunction $f^0$ from which $\O$ is computed. 
Or when $\r_x$ is zero at some site $x$,
because then the projection to U(1) or SU(2) is impossible.
The subspace of these exceptional configurations has codimension one.
Besides this, the global phase of
an eigenfunction is arbitrary and for SU(2) the two degenerate
eigenfunctions can be rotated by an arbitrary SU(2) transformation. 
This  implies that $\O$ in eqs. (\ref{DEFOMU}) and (\ref{DEFOMSU}) 
is only defined up to a global U(1) or SU(2) factor.
This global gauge transformation, however, is easy to fix by an (arbitrary)
prescription, e.g. $\O_{x_0}=1$ at a given site $x_0$.

Also after fixing the global transformation, there is  still the
subspace of codimension one where the gauge is not determined, 
which we shall refer to as the Gribov horizon  of the Laplacian gauge. 
Since it has measure zero in the gauge field configuration space,
it can be excluded from the  integration region of the path
integral. This can be implemented by prescribing
that the action is equal to infinity ($e^{-S}=0$) on the Gribov horizons.

In practice the gauge is not well defined when the lowest two eigenvalues 
differ by less than some small number $\e$, which is fixed by the numerical
precision that can be obtained in computing the eigenvalues. Similarly the
second kind of ambiguity arises in practice when 
$\r_x$ at some site $x$, is smaller than another threshold 
$\e'=O(\e)$.  The prescription to give infinite action to configurations 
on the Gribov horizon then amounts to excluding a small region of the
configurations space which no longer has measure zero but extends $O(\e)$ 
around the horizons. 
For finite $\e$ the exclusion of the small region around the
Gribov horizons will introduce a small systematic error of order
$\e$, but this effect should be very small ($\e$ will be typically $\ll
10^{-6}$) and can be controlled by
increasing the numerical precision with which the lowest eigenvalue and
eigenfunction of the Laplacian are computed.

To implement Laplacian gauge fixing numerically,
we first compute the smallest eigenvalues of the Laplacian
with a Lanczos algorithm. Since  only the smallest 
one or two eigenvalues are needed, we only have to iterate until the
desired eigenvalues have converged, which makes this
algorithm very efficient.  Using the, usually very
accurate, estimate of the smallest eigenvalue provided by the Lanczos
algorithm, we apply inverse iteration to compute the corresponding
eigenfunction and further improve the accuracy of the eigenvalue.
If more eigenfunctions are required, we proceed with the inverse
iteration, while projecting out the eigenfunction(s) found previously.
In this way we can typically compute the eigenvalues and eigenfunctions
to a precision $\d < 10^{-10}$, where $\d = ||\l^0f^0 - \D f^0||/||f^0||$.

To get some idea how frequently Gribov horizons are crossed in an actual 
numerical simulation and to test the numerical stability of the
algorithm, we have computed the lowest two eigenvalues of the Laplacian
and the corresponding eigenfunctions
for a sequence of gauge fields produced in a simulation of the pure
gauge model.
Since the exceptional configurations described above,
will be very rare, it is only
likely that we find one in an equilibrium ensemble, if this ensemble is
extremely large. On the other hand, when using a Monte Carlo
simulation to generate such an ensemble, subsequent configurations are
correlated. For instance  in a Hybrid Monte Carlo (HMC) simulation
subsequent configurations differ by $O(dt)$, with $dt$ the
time step in the HMC evolution.  Such a sequence of gradually changing
gauge field configurations follows a one dimensional curve in
configuration space and since the Gribov horizons have codimension one,
it appears likely that these horizons will be encountered along the way.
Hence, following the flow of the lowest eigenvalues of the
Laplacian for such a sequence of gauge fields, should be a good strategy 
to try to find exceptional configurations.

We have used a HMC algorithm to produce a chain of U(1) fields in two 
dimensions and a Metropolis algorithm to get a sequence of SU(2) fields 
in four dimensions. The results for SU(2) on a $8^4$ lattice at
$\b=2.0$ are shown in fig. 1. We are using a rather small value for
$\b$ in order to increase the fluctuations in the values of the smallest
eigenvalues of the Laplacian and to enhance the chance of observing a
level crossing.
The ragged lines represent the lowest
two eigenvalues (which are each twofold degenerate) of the Laplacian.
Subsequent gauge fields are separated by a single two-hit Metropolis sweep 
with a maximum update angle of $0.05\pi$.

\begin{figure} [ttt]
 \centerline{ \epsfysize= 9.5cm   \epsfbox{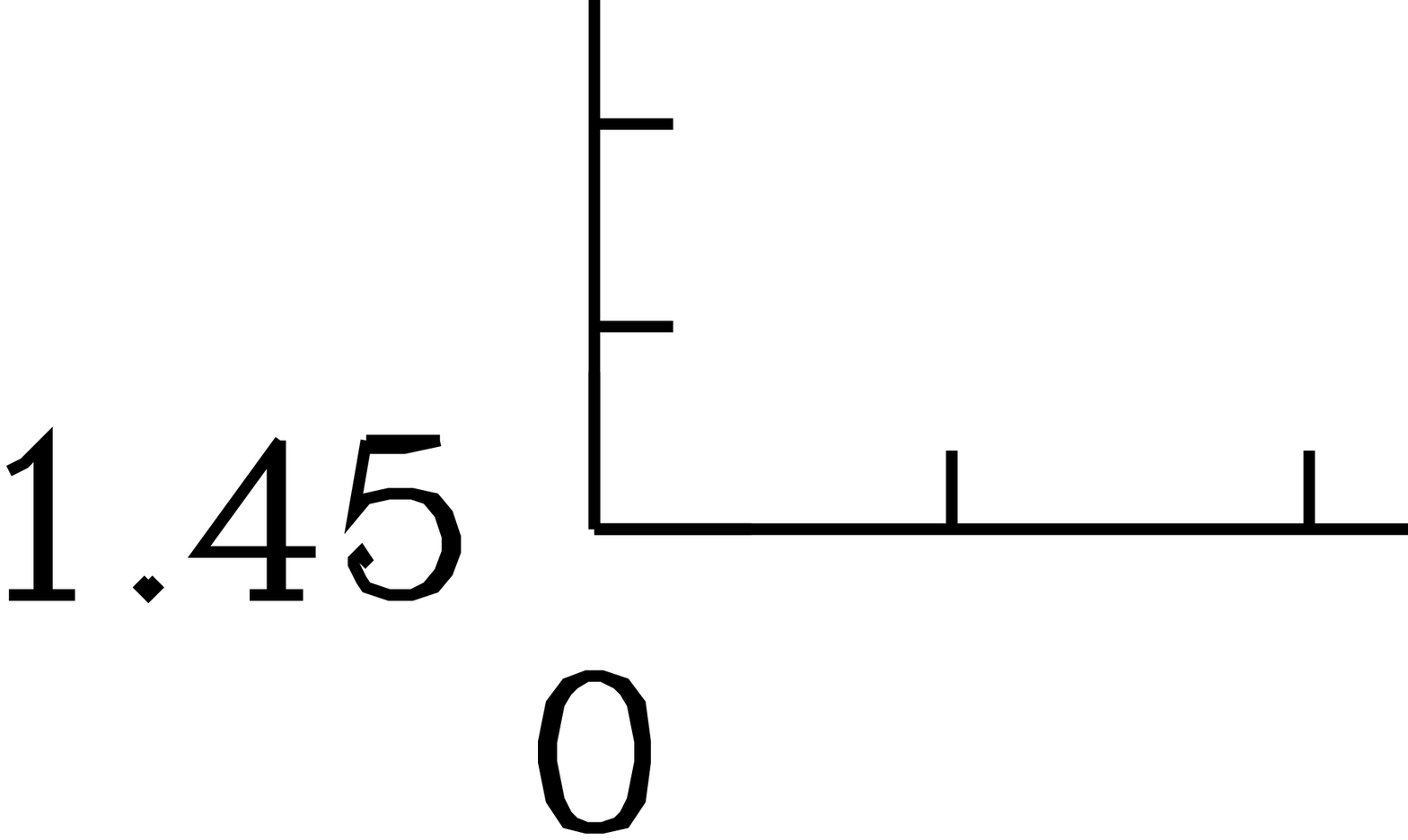}
 }
\capit{  Lowest two (non-degenerate) eigenvalues of the Laplacian
for SU(2) gauge fields along a Markov chain which is produced with
a Metropolis algorithm. Subsequent configurations are separated by
one update sweep (with two hits per site) of all lattice links, with a 
maximum update angel of $0.05\pi$; the lattice size is $8^4$ and $\b=2.0$.
}
\end{figure}

One sees that the two eigenvalues fluctuate considerably,
but they never  come very close to each other.
At the points where they come closets to each other, their separation is
still many order of magnitude larger than the precision with which the
eigenvalues can be computed. 
We also computed the eigenvalue flow at the larger value of $\b=2.5$
inside the scaling region.
Here we found that the smallest two eigenvalues on the average differ by
0.15, whereas their typical fluctuations are much smaller, $\approx
0.03$. Therefore level crossings are almost excluded in this case.

A similar picture emerges when looking at
the eigenvalue flow for U(1) gauge fields in two dimensions, as shown in
fig. 2. The lattice size is $16^2$ and $\b=2$.
Here the eigenvalues flow much more smoothly, because we are
using an HMC algorithm. The evolution of these gauge fields is governed by 
classical equations of motion except at the occasional momentum
refreshments.
We have taken the lattice volume unnecessarily large for the $\b$
considered (the string tension correlation length is $\x\approx
(2\b)^{1/2}$) in order to make the typical separation between the 
eigenvalues smaller than their fluctuations.

\begin{figure} [ttt]
 \centerline{ \epsfysize= 9.5cm   \epsfbox{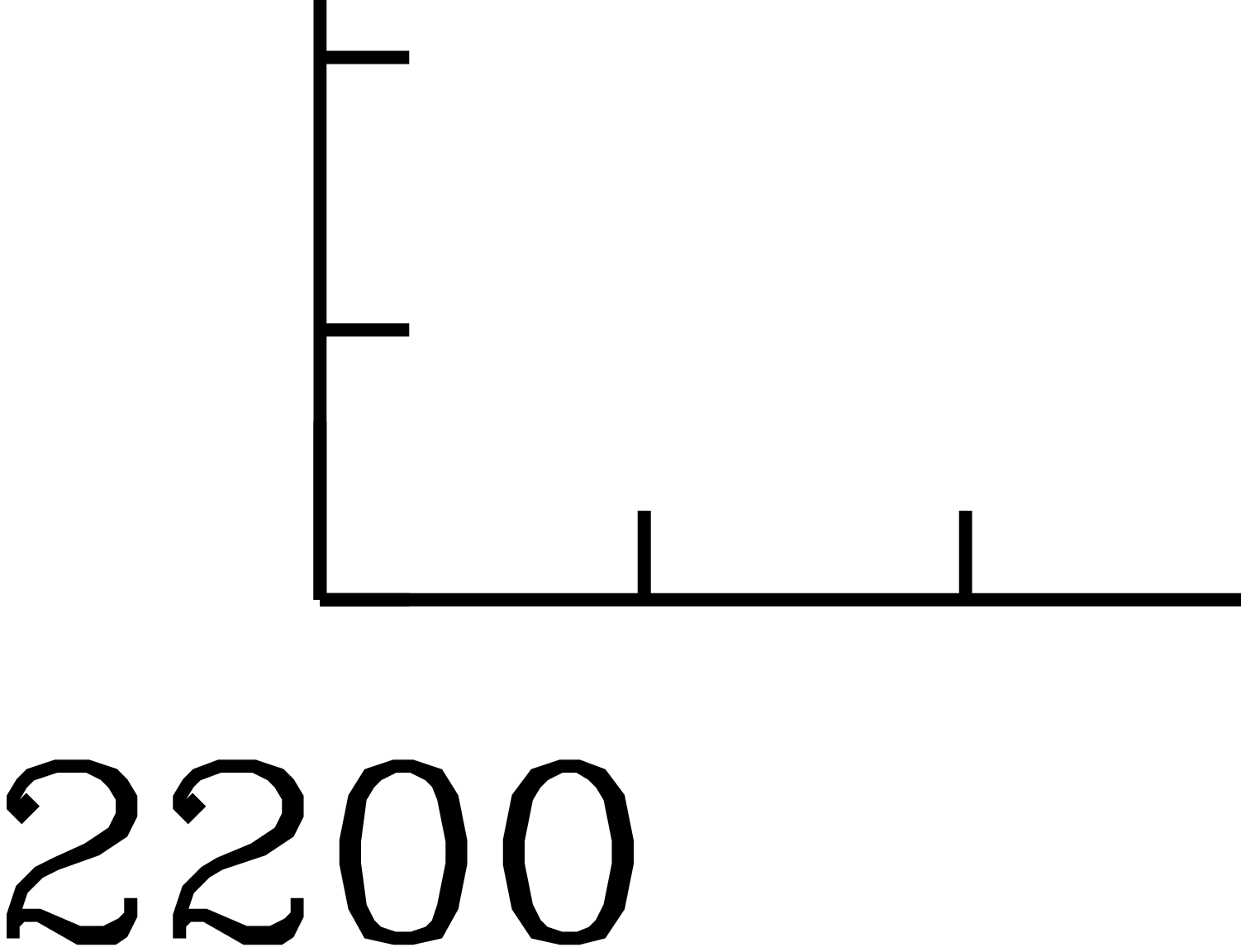}
 }
\capit{  Lowest two eigenvalues of the Laplacian
for U(1) gauge fields along a Markov chain which is produced with
a Hybrid Monte Carlo algorithm. Subsequent configurations are separated by
one HMC time step $dt=0.1$ and a momentum refreshment every 10 steps.
The lattice size is $16^2$ and $\b=2.0$.
}
\end{figure}

For larger $\beta$ at the same lattice volume $V=N^d$, the fluctuations of the
eigenvalues eventually  become smaller than the typical separation of the 
eigenvalues, which is of order
$2\pi/N$, and level crossings become increasingly unlikely.
On the other hand, going deeper into the scaling region for $\b\ra \infty$ 
one would
increase $N$ such that the physical lattice size $L=Na(\b)$ stays finite
with $a(\b)$ the lattice distance. Then one expects that the typical
separation between eigenvalues of the Laplacian decreases $\propto a$,
whereas the fluctuations of these eigenvalues are expected to be of
order $1/\b$. In four dimensional gauge theories this implies that the
fluctuations of the eigenvalues will become much larger than their typical
separation.

If the results of figs. 1 and 2, which suggest that the eigenvalues avoid
level crossing, remains valid also deeper into the scaling region where
the eigenvalue levels are much closer to each other, 
it should still be numerically feasible to compute the lowest eigenvalues 
and eigenfunctions also for large $\b$ on correspondingly large
lattices. However, it has to be expected that the high density of
small eigenvalues on increasingly large lattices, at some point defeats 
attempts to compute them with sufficiently high accuracy.

In order now to implement the prescription discussed in the previous
subsection, that configurations on the Gribov horizon must be excluded
form the path integral, we would have to compute the lowest two eigenvalues
for each configuration on the HMC trajectory, as in figs. 1-2. Then it
can be checked when these eigenvalues come so close to each other that
the configuration has to be considered as lying on the Gribov horizon.
Restricting one self to lattice sizes $N\aplt 32$,
the test runs suggest that the gauge fields 
that are important for a stochastic sampling of the path integral, 
on lattice sizes and at values of the gauge coupling that are presently used,
tend to avoid the Gribov horizons. 
On a Markov chain with 4000  U(1) configurations, we only found a single
configuration for which the two smallest eigenvalues were closer than
0.002 apart (at $t/dt=2393$ in fig. 2 we found $\l_0 = 0.2893332$ and
$\l_1 = 0.2904288$). 
Also for this nearest degeneracy, however, the two smallest
eigenvalues and the corresponding eigenfunctions could easily be computed 
without ambiguity.

\section{Comparison with the Landau gauge}

The objective of using gauge fixed gauge fields for the computation of
e.g. quark propagators, extended quark bilinears without a string of
$U$-fields in between or a chiral fermion determinant
is to remove the high-momentum modes of
the gauge fields which are present because of the gauge freedom. 
In this way, the situation in perturbative calculations can be
mimicked, where the propagator, fermion bilinears or the chiral
fermion determinant are calculated for external
gauge fields with momenta that are low compared to the fermion cutoff.
The Landau gauge then is a convenient gauge that leads to smooth gauge fixed 
fields. In this gauge one maximizes the value of the average link,
\be
  H = \RE \sum_{x\m} \Tr U_{\m x}. \label{LANF}
\ee
At an extremum, $H$ is stable under an infinitesimal gauge
transformation $U_{\m x} \ra (1+i\o_x)U_{\m x}(1-i\o_{x+\mh})$, which
implies that
\be
  \sum_\m \del^\pr_\mu\, \IM \,U_{\m x} = 0,
\ee
where the backward lattice derivative is defined by 
$\del^\pr_\m f_x = (f_x - f_{x-\mh})/a$.
Writing $U_{\mu x} = e^{iagA_{\m x}}$, this corresponds to the
Landau condition $\del^\pr_\m A_{\m x}=0$ up to lattice artifacts of 
$O(a^2)$.  

For the Laplacian gauge it is not immediately clear that the gauge
fixed configurations are equally smooth as in the Landau gauge.
To see in what way  the Laplacian gauge is related to the
Landau gauge, we can rewrite the Landau condition of maximizing
$H$ as a minimization of $Q$ defined as,
\be
 Q = \RE \sum_{x\m}\Tr \left( 2\O_x\O_x^\dg  
          - \O_xU_{\m x}\O_{x+\mh}^\dg
          - \O_{x+\mh} U^\dg_{\m x}\O_x^\dg \right) \label{MIN}
\ee
with $\O\in$U(1). Finding the absolute minimum is difficult because 
$\O$ is constraint to lay in the gauge group.  For non-trivial
gauge fields, $U\not=1$, this minimization is similar to finding the
ground state of a spin glass. In practice one can find a local minimum
easily but finding the absolute minimum is a `noncomputable' problem.
By relaxing the constraint that $\O\in$U(1) and replacing
$\O_x$ by $\r_x\O_x$, with $\r_x>0$ and $\sum_x \r_x^2=1$
one recognizes that the minimization (\ref{MIN}) turns into minimizing
the quadratic form $\sum _{xy} f^*_x\D_{xy}f_y$, with $\D$ the Laplacian
defined in (\ref{LAPL}) and $f_x=\r_x\O_x$. The solution is now
easily found and given by the eigenfunction
of the Laplacian with the smallest eigenvalue. This is easily seen for
$G=$U(1), but also holds for $G=$SU(2). 
This illustrates that 
the difference between the Landau and the Laplacian gauge lies in the
the `weight' function $\r_x$. In the Landau gauge  $\r\equiv 1$, 
whereas in the Laplacian gauge it is allowed to deviate from one
\cite{ViWi92}.

This can be made more explicit by using perturbation theory, writing
$U^\O_{\m x}= e^{iagA_{\m x}}$ and expanding in $agA$. Such an expansion
should be reasonable for the gauge fixed field $U^\O$.
To lowest order one finds that $\r_x = \r^{(0)} = V^{-1/2}$ is constant, 
with $V$ the lattice volume. The lowest order correction $\propto gA$
can be computed in perturbation theory and is found
to vanish. This implies that we can write
$\r_x = \r^{(0)} + g^2\r^{(2)}_x + O(g^3,a)$.
For  $U$ in the Laplacian gauge $Q =\sum _{xy} \r_x\D_{xy}(U)\r_y$  is
stable under an infinitesimal gauge transformation which implies the
differential gauge condition,
\be
 \sum_\m \del^\pr_\m A_{\m x}  = 
     -2 \sum_\m \frac{\del^\pr_\m \r_x}{\r_x} A_{\m x} + O(a).
	        \label{LAPDIF}
\ee
For smooth functions $\r_x$ with $\del_\m\r/\r\ll 1$, the rhs is
close to zero and
condition (\ref{LAPDIF}) approximately reproduces the differential 
Landau condition $\del^\pr_\m A_\m = 0$.
Using the perturbative result mentioned above, one sees
that $\del^\pr_\m\r_x/\r_x = g^2\del^\pr_\m\r^{(2)}_x/\r^{(0)} +
O(a,g^3)$, which vanishes $\propto g^2$ for $g\ra 0$.
On the other hand, for an almost exceptional configuration
where $\r_{x_0}\approx 0$, eq. (\ref{LAPDIF}) suggests that the Laplacian
gauge may deviate substantially form the Landau gauge in the lattice region
near the site $x_0$.

Since it is important to establish that the Laplacian gauge leads to
equally smooth gauge fixed fields as the Landau gauge also outside the
perturbative regime, we shall also compare the two gauges numerically.

\begin{figure} [ttt]
 \centerline{ \epsfysize= 9.5cm   \epsfbox{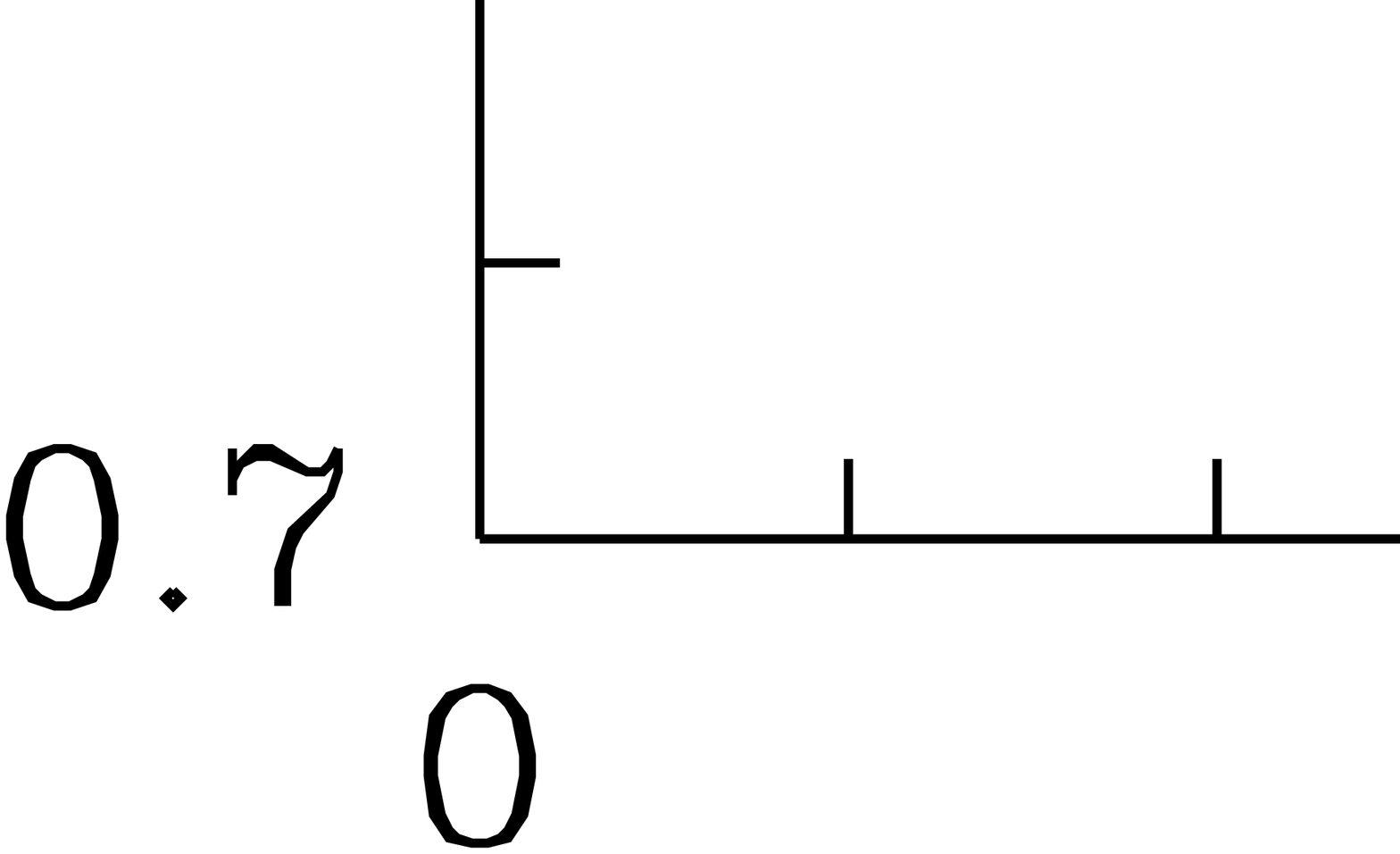}
 }
\capit{  Average link $\av{U}$ as a function of the inverse gauge 
coupling $\b$ for U(1) gauge fields in two dimensions, after standard 
Landau gauge fixing
(solid line), Laplacian gauge fixing (dashed line) and Laplacian
followed by Landau gauge fixing (dotted line). 
The lattice size is $20^2$.
}
\end{figure}

An obvious quantity to compare is the average link of the gauge fixed
gauge field. In fig. 3 the average link 
$\av{U}=\sum_{\m x} \RE \av{U_{\m x}}/V$
is shown for two dimensional U(1) gauge fields, as a function of the
inverse gauge coupling $\b$.
If $\av{U}$ is close to one, the configuration should be smooth. 
The dashed line is the result after
Laplacian gauge fixing, the full line is the result after standard
Landau fixing. Here we use a checker board relaxation algorithm to
maximize the function (\ref{LANF}). One sees, that the Laplacian
gauge fixing leads to gauge fixed configurations with a larger
average link for $\b \apgt 3$, which roughly corresponds to the 
scaling region of this model. Only for small $\b$ the usual Landau 
fixing produces a larger average link, but even there the difference
is not dramatic. The third curve (dots) is obtained by applying 
Landau gauge fixing {\em after} putting the gauge field in the Laplacian
gauge.  Here we consistently find that the subsequent Landau cooling
increases the average link to a somewhat larger value. 
The standard Landau gauge fixing algorithm is typically unable to find the
absolute maximum of (\ref{LANF}), and the lattice Landau gauge suffers
form Gribov ambiguities, for recent work on lattice Gribov copies,
see e.g. ref. \cite{GriLa}.  
After preconditioning with Laplacian
gauge fixing a larger maximum is found, but we have not
investigated if this preconditioning actually leads to the absolute maximum. 

Similarly we have computed the average link for SU(2) gauge fields in
four dimensions.  Here we also find little difference between the
Laplacian and Landau gauge for $\b\apgt 2.2$, see  fig.~5. To compute the
average link shown in fig.~5, we used 8 independent equilibrium
configurations and to see the presence of Gribov copies we applied  20 
different random gauge transformations to each of them.
Typically the average link depends on the initial random gauge transformation 
of the gauge field, but the differences in the final values of the
average link are usually small, less than $\approx 0.005$ for $\b\apgt 2$,
which would be invisible on the scale of fig. 4. 
For increasing $\b$ on a fixed lattice
size, the number of Gribov copies encountered, decreases and also
the difference in values of the average link in the various copies
appears to decrease. For smaller values of $\b$ the ambiguities
increase and also the difference between Landau and Laplacian
gauge fixed configurations becomes larger.

\begin{figure} [ttt]
 \centerline{ \epsfysize= 9.5cm   \epsfbox{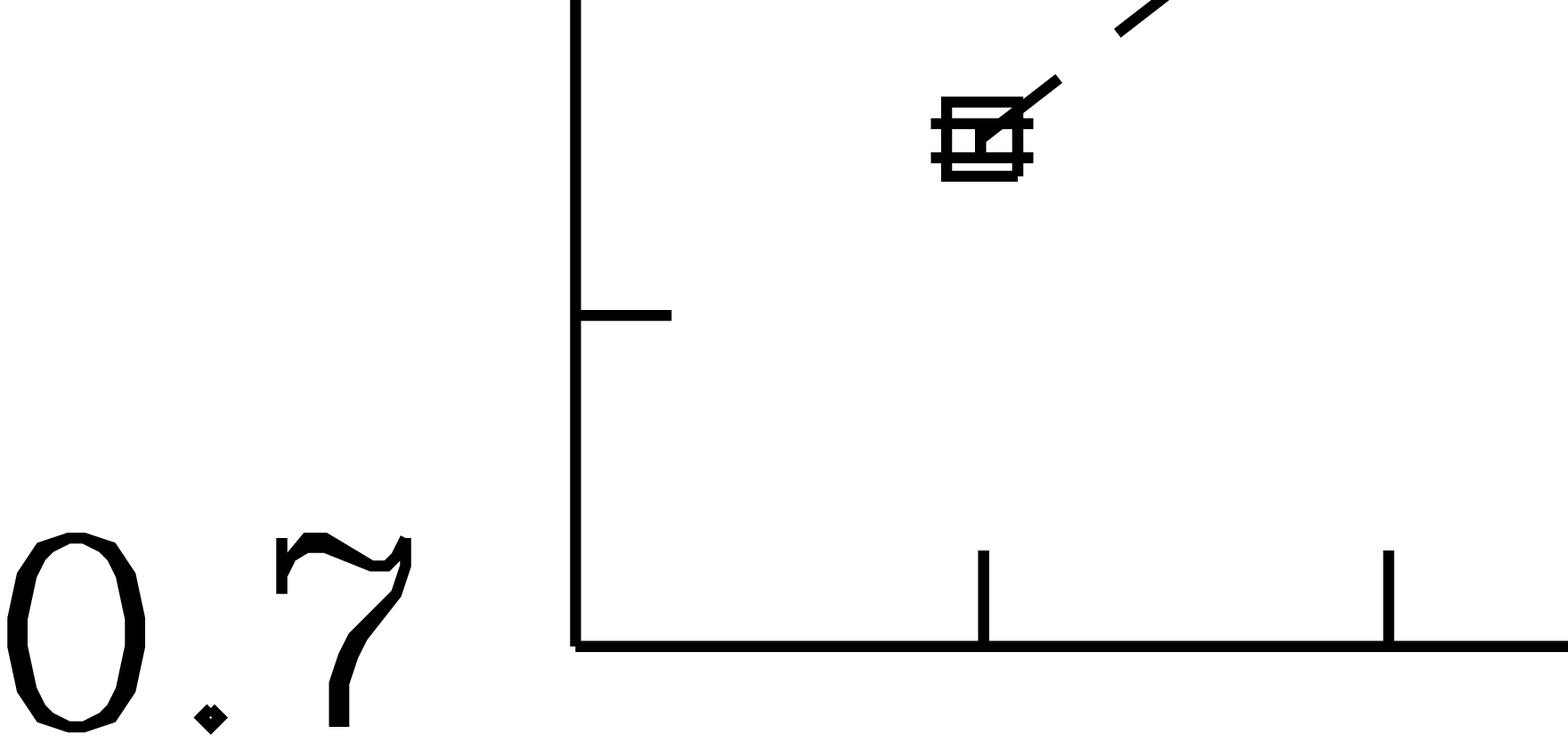}
 }
\capit{  Average link $\av{U}$ as a function of the inverse gauge 
coupling $\b$ for SU(2) gauge fiels in four dimensions, after standard 
Landau gauge fixing
(solid line) and Laplacian gauge fixing (dashed line).
The lattice size is $8^4$.
}
\end{figure}

The average link singles out the zero momentum mode of the trace of the
link field. As in perturbation theory one would like to see that also
the other small momentum modes of the gauge fixed field are boosted
compared to the high momentum modes. This can be illustrated by computing
the Fourier spectrum of the gauge fixed gauge field. In fig. 5 we plot the
average value of the coefficients $c_p$ of the momentum modes of the
SU(2) gauge fixed gauge field,
\be
  c_p^k = 
    \av{\left|\sum_{\m x} u^k_{\m x}e^{-ipx}/4V\right|^2}^{1/2},
\ee
where the $u^k$ are the real components of the SU(2) gauge field,
$U=i\sum_k^3\s_ku^k + u^4$.

\begin{figure} [ttt]
 \centerline{ \epsfysize= 9.5cm   \epsfbox{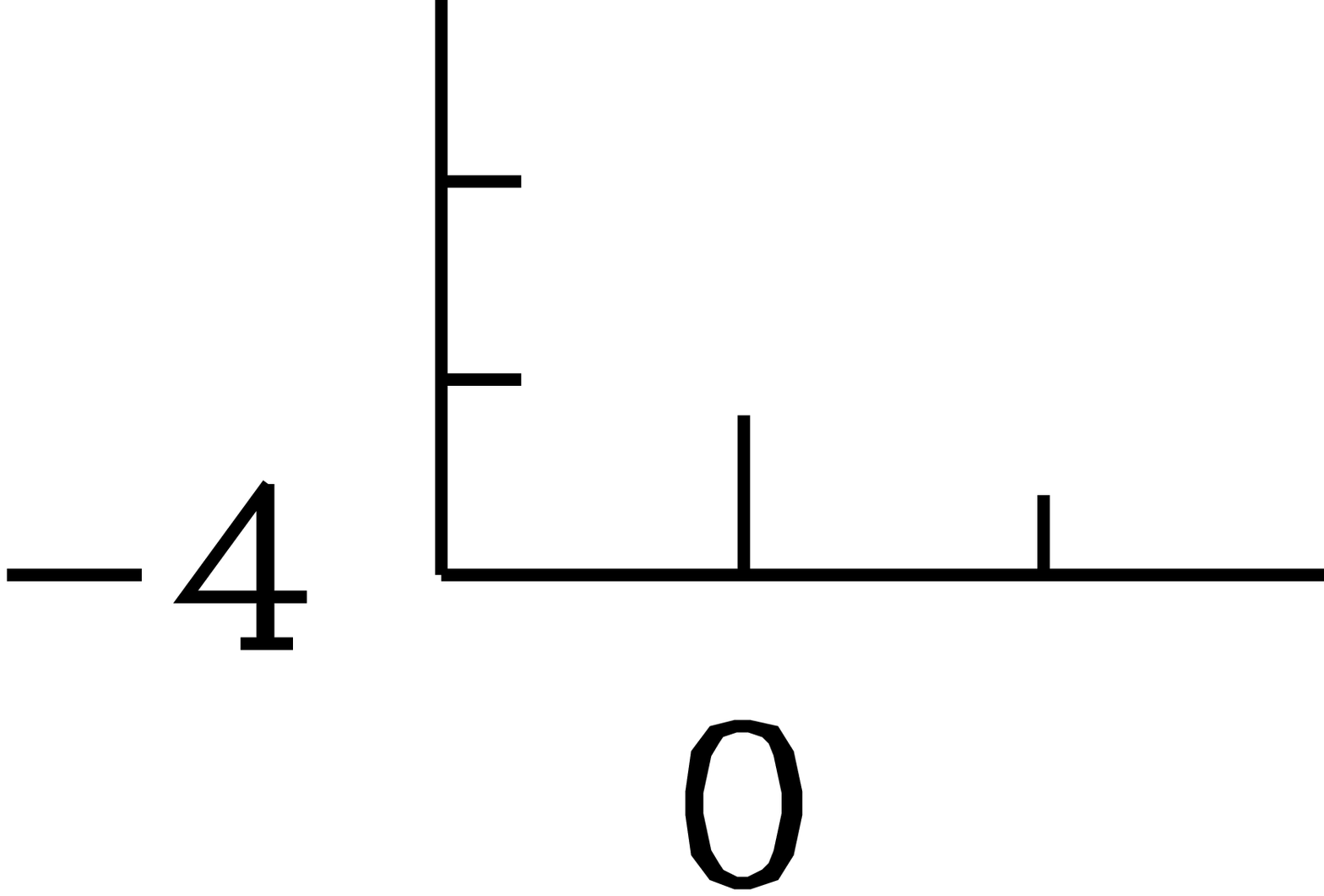}
 }
\capit{ Mean square average of the momentum modes of equilibrium SU(2) gauge 
fields at $\b=2.4$ on a $16^4$ lattice. The average $u^4$ component $c_p^4$ 
is shown without gauge fixing (open circles), after Laplacian gauge
fixing (boxes) and after Landau gauge fixing (crosses). Similarly the $c_p^1$
component is shown without gauge fixing (stars), after Laplacian gauge
fixing (triangles) and after Landau gauge fixing (three prongs). The
results for $c_p^2$ and $c_p^3$ are almost indistinguishable from that 
for $c_p^1$.
To avoid cluttering, we have suppressed some of the momenta for
$\ph^2>2$.
}
\end{figure}

The modes in fig. 5 are labeled by the value of
the lattice momentum $\ph=(\sum_\m (2-2\cos p_\m))^{1/2}$; the lattice
size is $16^4$ and we used 10 equilibrium configurations at $\b=2.4$ for
the average.
The zero momentum mode $c_0^4$ is responsible for the large value of the
expectation value of the average link, which
is shown already in fig.~5.
One sees that the $u^k$ components with $k\not=4$, which represent the
three components of the SU(2) gauge potential, are boosted
for small momenta $\ph^2<2$. The large momentum modes, $\ph^2\apgt 2$
of these components as well as all nonzero modes of $u^4$
are suppressed, as expected. 
This is particularly clear in
comparison with the result for non-gauge fixed fields. Here it is
seen that all momentum components are equally important  and the
$k\not=4$ modes are indistinguishable from the $k=4$ mode (open
circles and stars in fig. 5). 
Fig. 5 contains both the result for the Landau gauge
(open boxes and triangles) and for the Laplacian gauge (crosses and
three prongs). It shows that the
relative importance of the momentum modes of the gauge fixed field 
is almost identical in both gauges, over the full range of lattice 
momenta. For smaller values of $\b$ this close agreement becomes less.

\section{Discussion}

In this paper we have performed various tests of the Laplacian gauge
fixing prescription.  Two important properties claimed in ref.
\cite{ViWi92} of this gauge are
that it is unambiguously computable for almost all gauge fields and that
it leads to similarly smooth gauge fixed fields as the Landau gauge.

The Gribov horizons of the Laplacian gauge have measure zero in the
gauge field configuration space in the ideal case that the hypersurface
where the lowest eigenvalues of the Laplacian cross, or where the
eigenvalue on some site vanishes, can be exactly computed. In practice
the numerical accuracy $\e$ with which the eigenfunction can be
computed, gives the horizons a volume $\propto \e$ and
we define the path integral by excluding these Gribov horizon
regions. We find that the numerical accuracy of our algorithm is high,
leading to a very small $\e$ of order $10^{-10}$.

We have attempted to find how frequently  Gribov horizons are
encountered in an actual simulation. Using a HMC simulation of 
U(1) and a Metropolis simulation of SU(2) gauge theory we produced
(highly correlated) Markov chains of gauge fields configurations. We never 
found that the configurations on these chains came within this $\e$ region 
around the horizons. 
Only very rarely the two smallest eigenvalues approached each other
sufficiently closely that it is likely that the Markov chain
actually crossed a Gribov horizon. This suggests that the probability
for gauge fields, and hence their ground state wave function, in the regions
around the Gribov horizons is small in the cases we have investigated.
It is stressed, however, that even if the chain crosses the
Gribov horizon, this presents no problem for the gauge fixing algorithm,
unless a configurations would accidentally land inside the $\e$ region
around the Gribov horizon.

To test the smoothness of the Laplacian gauge fixed configurations, we
first showed that for vanishing gauge coupling, $g\ra 0$, the Laplacian
gauge reduces to the Landau gauge. We have studied Laplacian gauge
fixing numerically  for U(1) and SU(2) gauge fields. We find very similar
results for the average link and for the relative importance of the various
non-zero momentum modes of the Landau and Laplacian gauge fixed configurations.
Further more, we find that the 
Laplacian gauge fixing procedure is rather efficient. When comparing  with
standard Landau gauge fixing (where we maximize $H$ until its relative
change is less than $10^{-12}$) we find that Laplacian gauge fixing typically
takes 1-2 times  {\em less} computer time. These favorable results of 
Laplacian gauge fixing suggest that, at least on presently used lattice 
sizes, it is a viable alternative for Landau gauge
fixing that could be used to avoid the Gribov ambiguities that afflict
the Landau gauge. It would therefore be interesting to repeat e.g. a 
calculation of glue ball masses with gauge variant glueball operators using 
the Laplacian gauge and compare the results with those obtained with the
Landau gauge.

A disadvantage of Laplacian gauge fixing is that it cannot easily be
implemented in perturbation theory. However, this difficulty can perhaps
be circumvented by using a Monte Carlo simulation to compute the 
various two and three point Green functions that are
needed to determine e.g. current renormalizations. It may be
possible to compute the $\b$ dependence of these Green functions sufficiently
accurately for a range of large values of $\beta$ deep in the perturbative 
regime,  and fit the results to a power series in $1/\b$. 
In such a numerical simulation Laplacian gauge fixing could 
easily be implemented.

\subsubsection*{Acknowledgement}

I would like to thank W. Bock, M. Golterman, J. Hetrick,
J. Kuti, S. Sharpe, J. Smit and P. van Baal for discussions.
This work is supported by the DOE under grant DE-FG03-91ER40546
and by the TNLRC under grant RGFY93-206.

\end{document}